\documentstyle[amssymb,preprint,prb,aps]{revtex}
\begin{document}

\begin{titlepage}

\title
{Effects of arbitrarily directed field on spin phase oscillations in 
biaxial molecular magnets}

\author{Hui Hu\footnote {To whom 
the correspondence should be addressed.\\
Email: huhui@Phys.Tsinghua.edu.cn} }
\address{Department of Physics, 
Tsinghua University, Beijing 100084, P. R. China}
\author{ Jia-Lin Zhu}
\address{Department of Physics, 
Tsinghua University, Beijing 100084, P. R. China\\
Center for Advanced Study, Tsinghua University,
Beijing 100084, P. R. China}
\author{Rong L\"{u}}
\address{Center for Advanced Study, Tsinghua University,
Beijing 100084, P. R. China}
\author{Jia-Jiong Xiong}
\address{Department of Physics, 
Tsinghua University, Beijing 100084, P. R. China}

\maketitle
\begin{abstract}
Quantum phase interference and spin-parity effects are  studied in biaxial 
molecular magnets in a magnetic field at an arbitrarily directed angle.
The calculations of the ground-state tunnel splitting are performed on the basis of 
the instanton technique in the spin-coherent-state path-integral representation, 
and complemented by exactly numerical diagonalization.
Both the Wentzel-Kramers-Brillouin exponent and the preexponential factor
are obtained for the entire region of the direction of the field.
Our results show that the tunnel splitting oscillates with the field for the small field angle, 
while for the large field angle the oscillation is completely suppressed. This distinct angular 
dependence, together with the dependence of the tunnel splitting on the field strengh, 
provide an independent test for spin-parity effects in biaxial molecular magnets.
The analytical results for the molecular Fe$_{8}$ magnet, are found to be 
in good areement with the numerical simulations, which suggests that even 
the molecular magnet with total spin $S=10$ is large enough to be treated as 
a giant spin system.

\noindent
{\bf PACS number(s)}:  03.65.Bz, 75.45+j, 75.50.Xx
\end{abstract}

\end{titlepage}

\section{Introduction}

In recent years, owing mainly to the rapid advances both in new technologies
of miniaturization and in highly sensitive SQUID magnetometry, there have
been considerable theoretical and experimental studies carried out on the
nanometer-scale magnets which exhibit macroscopic quantum phenomena(MQP).%
\cite{overview} A number of nanometer-scale particles in the
superparamagnetic regime have been identified as candidates for the
observation of MQP such as the tunneling of the magnetization (or the
N\'{e}el vector) out of metastable potential minimum through the classically
impenetrable barrier to a stable one, i.e., macroscopic quantum tunneling
(MQT), or, more strikingly, macroscopic quantum coherence (MQC), where the
magnetization (or the N\'{e}el vector) coherently oscillates between
energetically degenerate easy directions over many periods. Up to now,
molecular magnets have been the most promising candidates to observe MQP
because they have well-defined structures and magnetic properties. The
system that have recently attracted much attention are the molecules $\left[ 
\text{Mn}_{12}\text{O}_{12}\text{(CH}_3\text{COO)}_{16}\text{(H}_2\text{O)}%
_4\right] $ (in short Mn$_{12}$Ac) \cite{Mn12Ac} and $\left[ \text{(tacn)}_6%
\text{Fe}_8\text{O}_2\text{(OH)}_{12}\right] ^{8+}$ (in short Fe$_8$), \cite
{Fe8} where tacn is a macrocyclic ligand triazacyclononance. The Mn$_{12}$Ac
molecule contains four Mn$^{4+}$ ($S=3/2$) ions in a central tretrahedron
surrounded by eight Mn$^{3+}$ ($S=2$) ions. Oxygen bridges allow
superexchange coupling among the Mn ions, and both high-field and ac
susceptometry experiments indicate a $S=10$ ground state, resulting from the
four inner Mn$^{4+}$ spins being paralleled to each other and the other
eight Mn$^{3+}$ also parallel with the two groups antiparallel to each
other. The magnetization relaxation experiments, the dynamic susceptibility
measurements, and the dynamic hysteresis experiments all indicate that the
thermally assisted, magnetic-field-tuned resonant coherently quantum
tunneling of magnetization occurs between spin states in a large number of
identical Mn$_{12}$Ac molecules. \cite{Mn12Ac} Magnetic measurements have
shown that Fe$_8$ also has a spin ground state $S=10$, which arises from
competing antiferromagnetic interactions between the eight $S=5/2$ iron
spins with six spins being parallel and antiparallel to the other two spins. 
\cite{Fe8}

One of the most striking effects in the magnetic MQP is that for some spin
systems with high symmetries, the tunneling behaviors of magnetization seem
sensitive to the parity of total spin of the magnet. It has been
theoretically demonstrated \cite{loss,delft} that the ground-state tunneling
level splitting is completely suppressed to zero for the half-integer total
spin ferromagnets with biaxial crystal symmetry in the absence of an
external magnetic field, resulting from the destructive interference of the
Berry phase or the Wess-Zumino, Chern-Simons term in the Euclidean action
between the symmetry-related tunneling paths connecting two classically
degenerate minima. Such destructive interference effect for half-integer
spins is known as the topological quenching. \cite{garg93} But for the
integer spins, the quantum interference between topologically different
tunneling paths is constructive, and therefore the ground-state tunneling
level splitting is nonzero. While such spin-parity effects are sometimes
related to Kramers degeneracy, they typically go beyond this theorem in
rather unexpected ways. \cite{loss,garg93,garg99,braun} One recent
experimental method based on the Landau-Zener model was developed by
Wernsdorfer and Sessoli \cite{wernsd} to measure the very small tunnel
splitting on the order of $10^{-8}$K in molecular Fe$_8$ magnets. They
observed a clear oscillation of the tunnel splitting as a function of the
magnetic field applied along the hard anisotropy axis, which is direct
evidence of the role of the topological spin phase (Berry phase) in the spin
dynamics of these molecules.

Motivated by the experiment on topological phase interference or spin-parity
effects in the molecule Fe$_8$, \cite{wernsd} in this paper we investigate
the resonant quantum tunneling of the magnetization vector in molecular
magnets with biaxial crystal symmetry in the presence of an external
magnetic field at an arbitrarily directed angle in the ZY plane. By applying
the instanton technique in the spin-coherent-state path-integral
representation, \cite{garg92} we calculate both the
Wentzel-Kramers-Brillouin (WKB) exponent and the preexponential factor in
the ground-state tunnel splitting. Our results show that for the small angle 
$\theta _H$ of the applied magnetic field, the ground-state tunnel splitting
oscillates with the field for both the integer and half-integer spins, and
the oscillation behavior for integer spins is significantly different from
that for half-integer spins. However, this oscillation is completely
suppressed for the large angle region. The distinct angular dependence,
together with the dependence of the ground-state tunnel splitting on the
strength of the external applied magnetic field, may provide an independent
experimental test for spin-parity effects in molecular magnets. It is noted
that the instanton technique is semiclassical in nature, i.e., valid for
large spins and in the continuum limit. Whether the instanton technique can
be applied in studying the spin dynamics in the molecular magnet with $S=10$
(such as Fe$_8$) is an open question. We study this problem with the help of
exact diagonalization calculation. Our results show that the analytical
calculation based on the instanton technique agrees excellently well with
the exact diagonalization calculation, which strongly suggests that the
molecular magnet with $S=10$ can be treated as a giant spin system.

\section{Model and Method}

The system of interest is a molecular magnet at a temperature well below its
anisotropy gap, which has the following Euclidean action in the
spin-coherent-state representation, \cite{garg92}

\begin{equation}
S_E(\theta ,\phi )=\int d\tau \left[ iS\left( \frac{d\phi }{d\tau }\right)
-iS\left( \frac{d\phi }{d\tau }\right) \cos \theta +E(\theta ,\phi )\right] ,
\label{action}
\end{equation}
where $S$ is the total spin of the molecular magnet. The polar angle $\theta 
$ and the azimuthal angle $\phi $ label the spin coherent state. $E(\theta
,\phi )$ is the total energy of the molecular magnet which includes the
magnetocrystalline anisotropy energy and the Zeeman energy when an external
magnetic field is applied.

It is noted that the Euclidean action is written in the north-pole gauge,
and the first two terms in Eq. (\ref{action}) define the Wess-Zumino or
Berry term which arises from the nonorthogonality of spin coherent states.
The Wess-Zumino term has a simple geometrical or topological interpretation.
For a closed path, this term equals $-iS$ times the area swept out on the
unit sphere between the path and the north pole. The first term in Eq. (\ref
{action}) is a total imaginary-time derivative, which has no effect on the
classical equation of motion for the magnetization vector, but yields the
boundary contribution to the Euclidean action. Loss et al. \cite{loss} and
von Delft and Henley \cite{delft} studied the physical effect of this total
derivative term, and they found that this term is crucial for the quantum
properties of the magnetic particle and makes the tunneling behaviors of
integer and half-integer spins strikingly different.

In the semiclassical limit, the dominant contribution to the Euclidean
transition amplitude comes from finite action solutions of the classical
equations of motion (instantons), which can be expressed as the following
equations in the spherical coordinate system, \cite{garg92} 
\begin{eqnarray}
iS\left( \frac{d\overline{\theta }}{d\tau }\right) \sin \overline{\theta }
&=&\frac{\partial E}{\partial \overline{\phi }},  \label{eom_a} \\
iS\left( \frac{d\overline{\phi }}{d\tau }\right) \sin \overline{\theta } &=&-%
\frac{\partial E}{\partial \overline{\theta }},  \label{eom_b}
\end{eqnarray}
where $\overline{\theta }$ and $\overline{\phi }$ denote the classical path.
Note that the Euclidean action Eq. (\ref{action}) describes the $\left(
1\oplus 1\right) -$dimensional dynamics in the Hamiltonian formulation with
canonical variables $\phi $ and $P_\phi =S(1-$cos$\theta )$.

According to the instanton technique in the spin-coherent-state
path-integral representation, the instanton's contribution to the tunneling
rate $\Gamma $ for MQT or the ground-state tunnel splitting $\Delta $ for
MQC (not including the geometric phase factor generated by the topological
term in the Euclidean action) is given by \cite{garg92} 
\begin{equation}
\Gamma \ (or\ \Delta )=p_0\omega _p\left( \frac{S_{cl}}{2\pi }\right)
^{1/2}e^{-S_{cl}},
\end{equation}
where $\omega _p$ is the small-angle precession or oscillation frequency in
the well, and $S_{cl}$ is the classical action or the WKB exponent which
minimizes the Euclidean action of Eq. (\ref{action}). The preexponential
factor $p_0$ originates from the quantum fluctuations about the classical
path, which can be evaluated by expanding the Euclidean action to second
order in the small fluctuations.

We describe the molecular magnet with biaxial crystal symmetry by the
standard Hamiltonian, \cite{garg93} 
\begin{equation}
{\cal H}=k_1\hat{S}_z^2+k_2\hat{S}_y^2,  \label{hami}
\end{equation}
where $k_1>k_2>0$ are proportional to the anisotropy coefficients, and we
take the easy, medium, and hard axes as {\bf x}, {\bf y}, and {\bf z},
respectively. If the magnetic field is applied in ZY plane, at an arbitrary
angle $0\leqslant \theta _H\leqslant 90^o$ with {\bf z}, the Hamiltonian
becomes \cite{garg99,wernsd} 
\begin{equation}
{\cal H}=k_1\hat{S}_z^2+k_2\hat{S}_y^2-g\mu _BH_z\hat{S}_z-g\mu _BH_y\hat{S}%
_y,  \label{hami_mag}
\end{equation}
where $g$ is the land\'{e} factor, and $\mu _B$ is the Bohr magneton. The
Zeeman energy term associated with the applied field $\vec{H}\equiv \left(
0,H_y,H_z\right) $ $\equiv \left( 0,H\sin \theta _H,H\cos \theta _H\right) $
is given in the last two terms of the Hamiltonian. If the field is below
some critical value $H_c\left( \theta _H\right) $ (to be computed), the
Hamiltonian Eq. (\ref{hami_mag}) has two degenerate minima, and therefore
the magnetization can resonate between these two directions, providing a
case of MQC. Under the spin-coherent-state and the imaginary time
representation, the $E(\theta ,\phi )$ term in the Euclidean action is given
by 
\begin{eqnarray}
E(\theta ,\phi ) &=&k_1S^2\cos ^2\theta +k_2S^2\sin ^2\theta \sin ^2\phi
-g\mu _BSH_z\cos \theta -g\mu _BSH_y\sin \theta \sin \phi   \nonumber \\
&=&K_1\cos ^2\theta +K_2\sin ^2\theta \sin ^2\phi -2K_1(H_z/H_a)\cos \theta
-2K_1(H_y/H_a)\sin \theta \sin \phi ,
\end{eqnarray}
with $K_1=k_1S^2$ and $K_2=k_2S^2$ being the transverse and longitudinal
anisotropy coefficients respectively, and $H_a=2K_1/g\mu _BS$ being the
anisotropy field. Introducing $\lambda =K_2/K_1$, $\cos \theta _0=H\cos
\theta _H/H_a$, and $\sin \theta _0\sin \phi _0=H\sin \theta _H/\lambda H_a$%
, the $E(\theta ,\phi )$ reduces to 
\begin{equation}
E(\theta ,\phi )=K_1\left( \cos \theta -\cos \theta _0\right) ^2+K_2\left(
\sin \theta \sin \phi -\sin \theta _0\sin \phi _0\right) ^2+E_0,
\label{energy}
\end{equation}
where $E_0$ is a constant that makes $E(\theta ,\phi )$ zero at the initial
state. It is clearly shown in Eq. (\ref{energy}) that the energy minima of
system are at $\theta _1=\theta _0$, $\phi _1=\phi _0$ and $\theta _2=\theta
_0$, $\phi _2=\pi -\phi _0$, and therefore there are two different instanton
paths of opposite windings around hard anisotropy axis. We denote them by
instanton $A:\phi =\phi _0\longrightarrow \phi =+\pi /2\longrightarrow \phi
=\pi -\phi _0$, and instanton $B:$ $\phi =\phi _0\longrightarrow \phi =-\pi
/2\longrightarrow \phi =\pi -\phi _0$.

The critical field at which the energy barrier disappears can be determined
by,

\begin{eqnarray}
\cos \theta _0 &=&\frac{H\cos \theta _H}{H_a}\leqslant 1,  \label{crit_a} \\
\sin \phi _0 &=&\frac{H\sin \theta _H}{\lambda H_a\left( 1-\frac{H^2\cos
^2\theta _H}{H_a^2}\right) ^{1/2}}\leqslant 1.  \label{crit_b}
\end{eqnarray}
Taking into account Eqs. (\ref{crit_a}) and (\ref{crit_b}), we obtain 
\begin{equation}
H_c=\frac{\lambda H_a}{\left( \sin ^2\theta _H+\lambda ^2\cos ^2\theta
_H\right) ^{1/2}}.  \label{crit_mag}
\end{equation}
For the special cases $\theta _H=0,\ \pi /2$, we have $H_c=H_a,\ \lambda H_a$%
, respectively. The dependence of $H_c/H_a$ on $\theta _H$ is plotted in
Fig. 1 for $\lambda =0.71$.

Now we investigate the tunneling behaviors of magnetization by applying the
instanton technique in the spin-coherent-state path-integral representation.
First of all, we must find out the classical path $\bar{\theta}$ and $\bar{%
\phi}$ that satisfies the boundary condition. Along the classical path, $E(%
\bar{\theta},\bar{\phi})$ is conserved, so that the relation between $\bar{%
\theta}(\tau )$ and $\bar{\phi}(\tau )$ can be found purely by using energy
conservation. \cite{garg93,garg99} After some algebra, we obtain 
\begin{equation}
\cos \bar{\theta}=\frac{\left( \cos \theta _0-\lambda ^{1/2}y\sin \bar{\phi}%
\right) \pm i\lambda ^{1/2}\left( x\sin \bar{\phi}-\sin \theta _0\sin \phi
_0\right) }{1-\lambda \sin ^2\bar{\phi}},  \label{relation}
\end{equation}
where $x$, $y$ are variables defined by 
\begin{equation}
x=%
%TCIMACRO{\QOVERD\{ \} {(a^2+b^2)^{1/2}+a}{2}}
%BeginExpansion
{(a^2+b^2)^{1/2}+a \overwithdelims\{\} 2}%
%EndExpansion
^{1/2}\qquad \text{and}\qquad y=%
%TCIMACRO{\QOVERD\{ \} {(a^2+b^2)^{1/2}-a}{2}}
%BeginExpansion
{(a^2+b^2)^{1/2}-a \overwithdelims\{\} 2}%
%EndExpansion
^{1/2},  \label{xy}
\end{equation}
with $a=1-\cos ^2\theta _0-\lambda \sin ^2\bar{\phi}+\lambda \sin ^2\theta
_0\sin ^2\phi _0$ and $b=2\lambda ^{1/2}\sin \theta _0\cos \theta _0\sin
\phi _0$. Here we assume that the condition $a$ $\geqslant 0$ is always
fulfilled for the small magnetic field $H$. The positive and negative sign
in Eq. (\ref{relation}) are corresponding to instanton and anti-instanton
solutions for path $A$ or $B$, respectively. For the low magnetic field ,
one can take $\bar{\phi}(\tau )$ to be entirely real (see appendix A for
detail). \cite{garg93} Then through the expression $S_{cl}=iS\int_{-\infty
}^{+\infty }(1-\cos (\bar{\theta}))(d\bar{\phi}/d\tau )d\tau $, the desired
Wentzel-Kramers-Brillouin (WKB) exponents (or classical actions) are found
to be 
\begin{eqnarray}
%TCIMACRO{\func{Im}}
%BeginExpansion
\mathop{\rm Im}%
%EndExpansion
(S_{cl}^A) &=&S(+\pi -2\phi _0)-\frac{2S\cos \theta _0}{(1-\lambda )^{1/2}}%
\left( \frac \pi 2-\arctan \left( (1-\lambda )^{1/2}\tan \phi _0\right)
\right)   \nonumber \\
&&+2\lambda ^{1/2}S\int_{\phi _0}^{\pi /2}\frac{y\sin \phi d\phi }{1-\lambda
\sin ^2\phi },  \nonumber \\
&&  \nonumber \\
%TCIMACRO{\func{Re}}
%BeginExpansion
\mathop{\rm Re}%
%EndExpansion
(S_{cl}^A) &=&2\lambda ^{1/2}S\left\{ \frac{\sin \theta _0\sin \phi _0}{%
(1-\lambda )^{1/2}}\left( -\frac \pi 2+\arctan \left( (1-\lambda )^{1/2}\tan
\phi _0\right) \right) \right.   \nonumber \\
&&\left. +\int_{\phi _0}^{\pi /2}\frac{x\sin \phi d\phi }{1-\lambda \sin
^2\phi }\right\}   \label{SclA}
\end{eqnarray}
for instanton $A$, and 
\begin{eqnarray}
%TCIMACRO{\func{Im}}
%BeginExpansion
\mathop{\rm Im}%
%EndExpansion
(S_{cl}^B) &=&S(-\pi -2\phi _0)+\frac{2S\cos \theta _0}{(1-\lambda )^{1/2}}%
\left( \frac \pi 2+\arctan \left( (1-\lambda )^{1/2}\tan \phi _0\right)
\right)   \nonumber \\
&&+2\lambda ^{1/2}S\int_{\phi _0}^{\pi /2}\frac{y\sin \phi d\phi }{1-\lambda
\sin ^2\phi },  \nonumber \\
&&  \nonumber \\
%TCIMACRO{\func{Re}}
%BeginExpansion
\mathop{\rm Re}%
%EndExpansion
(S_{cl}^B) &=&2\lambda ^{1/2}S\left\{ \frac{\sin \theta _0\sin \phi _0}{%
(1-\lambda )^{1/2}}\left( +\frac \pi 2+\arctan \left( (1-\lambda )^{1/2}\tan
\phi _0\right) \right) \right.   \nonumber \\
&&\left. +\int_{\phi _0}^{\pi /2}\frac{x\sin \phi d\phi }{1-\lambda \sin
^2\phi }\right\}   \label{SclB}
\end{eqnarray}
for instanton $B$. Therefore, the difference of the action of two tunneling
paths is 
\begin{eqnarray}
%TCIMACRO{\func{Im}}
%BeginExpansion
\mathop{\rm Im}%
%EndExpansion
(\Delta S_{cl}) &=&%
%TCIMACRO{\func{Im}}
%BeginExpansion
\mathop{\rm Im}%
%EndExpansion
(S_{cl}^B)-%
%TCIMACRO{\func{Im}}
%BeginExpansion
\mathop{\rm Im}%
%EndExpansion
(S_{cl}^A)=-2S\pi \left( 1-\frac{\cos \theta _0}{(1-\lambda )^{1/2}}\right)
\equiv -2\Phi (H),  \label{ImDS} \\
%TCIMACRO{\func{Re}}
%BeginExpansion
\mathop{\rm Re}%
%EndExpansion
(\Delta S_{cl}) &=&%
%TCIMACRO{\func{Re}}
%BeginExpansion
\mathop{\rm Re}%
%EndExpansion
(S_{cl}^B)-%
%TCIMACRO{\func{Re}}
%BeginExpansion
\mathop{\rm Re}%
%EndExpansion
(S_{cl}^A)=2S\pi \sin \theta _0\sin \phi _0\left( \frac \lambda {1-\lambda }%
\right) ^{1/2}.  \label{ReDS}
\end{eqnarray}
Note that $%
%TCIMACRO{\func{Im}}
%BeginExpansion
\mathop{\rm Im}%
%EndExpansion
(\Delta S_{cl})$ comes from the Berry phase $iS(1-\cos \bar{\theta})\left( d%
\bar{\phi}/d\tau \right) $, and leads to the oscillation of the ground-state
tunnel splitting with the magnetic field.

The preexponential factor in the tunnel splitting can be evaluated by taking
the asymptotic form of zero-mode $\left( d\overline{\phi }/d\tau \right) ,$ 
\cite{garg92} which can be deduced from the classical equations of motion
(see appendix A).

Finally, the ground-state tunnel splitting is found to be 
\begin{eqnarray}
\hbar \Delta &=&\hbar \Delta _0\left| 1+e^{2i\Phi (H)}e^{-%
%TCIMACRO{\func{Re} }
%BeginExpansion
\mathop{\rm Re}%
%EndExpansion
\Delta S_{cl}}\right|  \nonumber \\
&=&\hbar \Delta _0\left\{ \left( 1-e^{-%
%TCIMACRO{\func{Re} }
%BeginExpansion
\mathop{\rm Re}%
%EndExpansion
\Delta S_{cl}}\right) ^2+4e^{-%
%TCIMACRO{\func{Re} }
%BeginExpansion
\mathop{\rm Re}%
%EndExpansion
\Delta S_{cl}}\cos ^2\Phi (H)\right\} ^{1/2},  \label{split0}
\end{eqnarray}
with 
\begin{equation}
\hbar \Delta _0=c\omega _0^{3/2}\frac{4\sin \theta _0}{\left( \sin ^2\theta
_0-\lambda \right) ^{1/2}}\left( \frac{S^2}{2\pi K_1}\right) ^{1/2}\left( 
\frac{\sin ^2\theta _0}{\sin ^2\theta _0+\lambda \cos ^2\theta _0\sin ^2\phi
_0}\right) ^{1/2}e^{-%
%TCIMACRO{\func{Re} }
%BeginExpansion
\mathop{\rm Re}%
%EndExpansion
(S_{cl}^A)},  \label{split1}
\end{equation}
where $\omega _0=(2K_1/S)\lambda ^{1/2}\sin \theta _0\cos \phi _0$. The
dimensionless prefactor $c$ can often be of order 1 or so, and is therefore
relevant to the exact diagonalization calculation.

\section{Results and Discussion}

Before we discuss the Eq. (\ref{split0}), we note here that our model can be
directly related to the model describing the molecular Fe$_8$ magnet, \cite
{wernsd} 
\begin{equation}
{\cal H}=-D\hat{S}_z^2+E(\hat{S}_x^2-\hat{S}_y^2)+g\mu _BH_x^{\prime }\hat{S}%
_x+g\mu _BH_y^{\prime }\hat{S}_y,
\end{equation}
with $K_1=(D+E)S^2$, $K_2=(D-E)S^2$ and $H_z=H_x^{\prime }$, $%
H_y=-H_y^{\prime }$. According to the typical parameters of Fe$_8$, $D=0.275$%
K, $E=0.046$K and $g=2$, we obtain that $K_1=32.1$K, $K_2=22.9$K, $\lambda
=0.71$ and $H_a=4.77$T. These parameters are used throughout the whole
calculation.

First, we discuss the effects of arbitrarily directed field on the spin
phase oscillation. Our results show that the topological phase interference
or spin-parity effects depend on the direction of the magnetic field
significantly. From Eq (\ref{split0}), whether the ground-state tunnel
splitting oscillates with the field is determined by two factors, $\Phi (H)$
and $%
%TCIMACRO{\func{Re}}
%BeginExpansion
\mathop{\rm Re}%
%EndExpansion
(\Delta S_{cl})$. For a fixed field strength, i.e. $H\equiv H_S\equiv \left( 
\sqrt{H_y^2+H_z^2}\right) $, it brings two-fold effects to increase the
field angle $\theta _H$: ({\it i}) the dependence of $\Phi (H)$ on the field
strength is reduced. As $\theta _H$ increases up to $\frac \pi 2,$ $\Phi
(H)\equiv $ $\pi S\left( 1-\frac{H\cos \theta _H}{(1-\lambda )^{1/2}H_a}%
\right) $ gradually reduces to a constant $\pi S.$ ({\it ii}) The degree of
oscillation $\sigma $ which is defined as a ratio of the oscillation part $%
4e^{-%
%TCIMACRO{\func{Re} }
%BeginExpansion
\mathop{\rm Re}%
%EndExpansion
\Delta S_{cl}}$ to the non-oscillation part $\left( 1-e^{-%
%TCIMACRO{\func{Re} }
%BeginExpansion
\mathop{\rm Re}%
%EndExpansion
\Delta S_{cl}}\right) ^2$ in Eq. (\ref{split0}) decreases from $+\infty $ to
a small value.

For $\theta _H=0$ (the field is along the hard anisotropy axis), the
symmetry of two classical paths imposes $%
%TCIMACRO{\func{Re}}
%BeginExpansion
\mathop{\rm Re}%
%EndExpansion
(\Delta S_{cl})=0,$ then, the tunnel splitting $\Delta =2\Delta _0\left|
\cos \Phi (H)\right| $ oscillates with the field and thus vanish whenever 
\begin{equation}
\frac H{H_a}=(1-\lambda )^{1/2}\frac{(S-n-1/2)}S,
\end{equation}
where $n=0,1,2,...$ is an integer. \cite{garg93} For $\theta _H=\frac \pi 2$
(the field is along the medium anisotropy axis), on the other hand, $\Phi
(H) $ becomes a constant $\pi S$, and therefore the tunnel splitting
increases monotonically with the field. When the field is applied in ZY
plane with an arbitrarily angle $\theta _H$, one may expect to observe a
crossover for a certain field angle $\theta _H^c$. Choosing the magnitude of 
$\vec{H}$ to be the first value that makes cos$\Phi (H)=0$ and taking $%
\sigma =1$, we obtain 
\begin{equation}
\theta _H^c\approx \arctan \left( \frac{1.76\lambda ^{1/2}}\pi \right)
\end{equation}
For the molecular Fe$_8$ magnet, $\lambda =0.71$, we find $\theta
_H^c\approx 25^o$, which agrees well with the exact diagonalization
calculation (shown in Fig. 2a) and the experimental results obtained by
Wernsdorfer and Sessoli. \cite{wernsd} It is noted that the value of $\theta
_H^c$ is independent of the total spin $S$ of the molecular magnet, and
depends on the parameter $\lambda $ only. For highly anisotropic materials,
the typical values of the transverse and longitudinal anisotropy
coefficients are $K_1\approx 10^7$erg/cm$^3$ and $K_2\approx 10^5$erg/cm$^3$%
. Thus, $\lambda =0.01$, the value of $\theta _H^c$ is estimated to be $%
3.2^o $ and is much smaller than that for the molecular Fe$_8$ magnet, which
means that even a very small misalignment of the field with the hard
anisotropy axis can completely destroy the oscillation. Therefore, the
molecular Fe$_8$ magnet is a better candidate for observing the oscillation
of the ground-state tunnel splitting with the field compared with the highly
anisotropic materials.

\smallskip Next, we turn to the comparison between the analytical
calculations and the numerical stimulations. It is noted that the instanton
approach is semiclassical in nature, i.e., valid for large spins and in the
continuum limit. Whether the instanton technique can be applied in studying
the spin dynamics in molecular magnets with $S=10$ (such as Fe$_8$) is an
open question. We have performed the numerical diagonalization of the
Hamiltonian Eq. (\ref{hami_mag}) for the molecular Fe$_{8\text{ }}$magnet in
the presence of an external magnetic field at an arbitrarily directed angle
in ZY plane. As illustrated in Fig. 2a, for Fe$_8$, the analytical results
based on the instanton technique are in good agreement with the exact
diagonalization results. In order to show the agreement more clearly, we
have presented the numerical and analytical results in Table 1 for the
molecular Fe$_8$ magnet in the magnetic field applied at angle $\theta
_H=0^o,\ 15^o,\ 30^o$ and $90^o$. It is clearly shown that the accuracy of
the semiclassical calculation is very high for the low magnetic field up to $%
H=1.0$T for the entire region of the angle $0^o\leqslant \theta _H\leqslant
90^o$. As a result, we conclude that the molecular magnet with $S=10$ can be
treated as a giant spin system. The numerical and analytical calculated
tunnel splitting as a function of the field are shown in Fig. 2b for
half-integer spin $S=9.5$, and the good agreement between numerical and
analytical results is also found. From Fig. 2, it is obviously that the
tunneling behavior of magnetization of integer spins is significantly
different from that for half-integer spins. At the end of this section, we
present the results with different parameter $\lambda $ for the field along $%
{\bf z}$ or ${\bf y}$ in Figs. (3a) and (3b). It is interesting to note that
the tunnel splitting increases rapidly by lowing the parameter $\lambda $,
which suggests that highly anisotropic materials are more suitable for
observing MQP in experiments.

In conclusion, we have studied the ground-state tunnel splitting in the
molecular magnets with biaxial crystal symmetry in the presence of an
external magnetic field at an arbitrarily directed angle. The switching from
oscillation to the monotonic growth of the ground-state tunnel splitting on
the field angle has been shown in detail. Our results are suitable for a
quantitative description of some aspects of the new experimental behavior on
the molecular Fe$_8$ magnets. \cite{wernsd}

\section*{Acknowledgments}

The authors are indebted to Professor W. Wernsdorfer and Professor R.
Sessoli for providing their paper (Ref. 9). The financial support from
NSF-China (Grant No.19974019) and China's ''973'' program is gratefully
acknowledged.

\begin{center}
{\bf Appendix A: Evaluate the ground-state tunnel splitting\ \ }
\end{center}

In this appendix, the general scheme for calculation of the ground-state
tunnel splitting is presented, the main assumptions and approximations are
also outlined.

We start with the relation between $\theta (\tau )$ and $\phi (\tau )$
obtained from the energy conversion \cite{garg93} (see Eq. (\ref{relation}%
)). Only one instanton, say the + one, need be considered explicitly (from
now on, we drop all the bars and identify $\theta =\bar{\theta}(\tau )$ , $%
\phi =\bar{\phi}(\tau )$ for convenient): 
\begin{eqnarray}
\cos \theta  &=&\frac{\left( \cos \theta _0-\lambda ^{1/2}y\sin \phi \right)
+i\lambda ^{1/2}\left( x\sin \phi -\sin \theta _0\sin \phi _0\right) }{%
1-\lambda \sin ^2\phi },  \label{costheta} \\
&&  \nonumber \\
\sin \theta  &=&\frac{\left( x-\lambda \sin \theta _0\sin \phi _0\sin \phi
\right) -i\left( \lambda ^{1/2}\cos \theta _0\sin \phi -y\right) }{1-\lambda
\sin ^2\phi }.  \label{sintheta}
\end{eqnarray}
In order to determine the classical paths, we need another equation of
motion(see Eq. (\ref{eom_b})), 
\begin{eqnarray}
iS\sin \theta \dot{\phi} &=&-\frac{\partial E(\theta ,\phi )}{\partial
\theta }  \nonumber \\
&=&2K_1(\cos \theta -\cos \theta _0)\sin \theta -2K_2(\sin \theta \sin \phi
-\sin \theta _0\sin \phi _0)\cos \theta \sin \phi .  \label{fai0}
\end{eqnarray}
After dividing the factor $2K_1\sin \theta $ on both sides and substituting
Eqs. (\ref{costheta}) and (\ref{sintheta}), we rewrite Eq. (\ref{fai0}) as 
\begin{eqnarray}
i\frac S{2K_1}\dot{\phi} &=&-\lambda ^{1/2}y\sin \phi +i\lambda ^{1/2}\left(
x\sin \phi -\sin \theta _0\sin \phi _0\right) +  \nonumber \\
&&\lambda \sin \theta _0\sin \phi _0\sin \phi \frac{\left( \cos \theta
_0-\lambda ^{1/2}y\sin \phi \right) +i\lambda ^{1/2}\left( x\sin \phi -\sin
\theta _0\sin \phi _0\right) }{\left( x-\lambda \sin \theta _0\sin \phi
_0\sin \phi \right) -i\left( \lambda ^{1/2}\cos \theta _0\sin \phi -y\right) 
}  \nonumber \\
&=&%
%TCIMACRO{\func{Re}}
%BeginExpansion
\mathop{\rm Re}%
%EndExpansion
f(\phi )+i%
%TCIMACRO{\func{Im}}
%BeginExpansion
\mathop{\rm Im}%
%EndExpansion
f(\phi ),  \label{fai1}
\end{eqnarray}
where 
\begin{eqnarray}
%TCIMACRO{\func{Re}}
%BeginExpansion
\mathop{\rm Re}%
%EndExpansion
f(\phi ) &=&-\lambda ^{1/2}y\sin \phi +\lambda \sin \theta _0\sin \phi
_0\sin \phi   \nonumber \\
&&\times \frac{\left( 1-\lambda \sin ^2\phi \right) \left( x\cos \theta
_0-\lambda ^{1/2}\sin \theta _0\sin \phi _0\right) }{\left( x-\lambda \sin
\theta _0\sin \phi _0\sin \phi \right) ^2+\left( \lambda ^{1/2}\cos \theta
_0\sin \phi -y\right) ^2},  \label{ReFun} \\
&&  \nonumber \\
%TCIMACRO{\func{Im}}
%BeginExpansion
\mathop{\rm Im}%
%EndExpansion
f(\phi ) &=&\lambda ^{1/2}\left( x\sin \phi -\sin \theta _0\sin \phi
_0\right) +\lambda \sin \theta _0\sin \phi _0\sin \phi   \nonumber \\
&&\times \frac{%
\begin{array}{c}
\left[ \left( \cos \theta _0-\lambda ^{1/2}y\sin \phi \right) \left( \lambda
^{1/2}\cos \theta _0\sin \phi -y\right) +\lambda ^{1/2}\times \right.  \\ 
\left. \left( x\sin \phi -\sin \theta _0\sin \phi _0\right) \left( x-\lambda
\sin \theta _0\sin \phi _0\sin \phi \right) \right] 
\end{array}
}{\left( x-\lambda \sin \theta _0\sin \phi _0\sin \phi \right) ^2+\left(
\lambda ^{1/2}\cos \theta _0\sin \phi -y\right) ^2}.  \label{ImFun}
\end{eqnarray}
As shown in Eq. (\ref{fai1}) , if $%
%TCIMACRO{\func{Re}}
%BeginExpansion
\mathop{\rm Re}%
%EndExpansion
f(\phi $ $)$ is always zero for arbitrary $\phi $, one can then take $\phi
(\tau )$ to be entirely real. Unfortunately, it is not the case, $%
%TCIMACRO{\func{Re}}
%BeginExpansion
\mathop{\rm Re}%
%EndExpansion
f(\phi $ $)$ is not always zero {\it except} $H_z=0$ or $H_y=0$. Checking
the right-hand-side of Eq. (\ref{ReFun}) more carefully, we find that up to
the leading term, Eq. (\ref{ReFun}) can be rewritten as 
\begin{eqnarray}
\left| 
%TCIMACRO{\func{Re}}
%BeginExpansion
\mathop{\rm Re}%
%EndExpansion
f(\phi )\right|  &\approx &\frac{\lambda ^2\sin ^3\phi }{\left( 1-\lambda
\sin ^2\phi \right) ^{1/2}}\cos \theta _0\sin \phi _0  \nonumber \\
&\leqslant &\frac{\lambda ^2}{\left( 1-\lambda \right) ^{1/2}}\cos \theta
_0\sin \phi _0  \nonumber \\
&\approx &\frac \lambda {\left( 1-\lambda \right) ^{1/2}}\frac{H_zH_y}{H_a^2}
\nonumber \\
&=&\frac \lambda {2\left( 1-\lambda \right) ^{1/2}}\frac{H^2\sin 2\theta _H}{%
H_a^2}.  \label{ReFunAppr}
\end{eqnarray}
It is clearly shown from Eq. (\ref{ReFunAppr}) that the $%
%TCIMACRO{\func{Re}}
%BeginExpansion
\mathop{\rm Re}%
%EndExpansion
f(\phi )$ is almost two orders of magnitude smaller than $%
%TCIMACRO{\func{Im}}
%BeginExpansion
\mathop{\rm Im}%
%EndExpansion
f(\phi )$ for the low magnetic field(i.e., for $H/H_a\leqslant 1/5$). So it
is sufficient to drop the term $%
%TCIMACRO{\func{Re}}
%BeginExpansion
\mathop{\rm Re}%
%EndExpansion
f(\phi )$ in Eq. (\ref{fai1}) for the first order approximation and treat $%
\phi (\tau )$ as a real variable. Under this assumption, the classical
Euclidean action can be evaluated easily without resolving the analytical
solutions of $\theta (\tau )$ and $\phi (\tau )$, 
\begin{eqnarray}
S_{cl} &=&iS\int_{-\infty }^{+\infty }(1-\cos (\theta (\tau )))(d\phi /d\tau
)d\tau   \nonumber \\
&=&iS\int_{\phi _i}^{\phi _f}(1-\cos (\theta (\phi )))d\phi   \nonumber \\
&=&iS\int_{\phi _i}^{\phi _f}\left( 1-\frac{\left( \cos \theta _0-\lambda
^{1/2}y\sin \phi \right) +i\lambda ^{1/2}\left( x\sin \phi -\sin \theta
_0\sin \phi _0\right) }{1-\lambda \sin ^2\phi }\right) d\phi ,  \label{Scl}
\end{eqnarray}
where $\phi _i=\phi (\tau \rightarrow -\infty )$, $\phi _f=\phi (\tau
\rightarrow +\infty )$. The results for instantons $A$ and $B$ are given in
Eqs. (\ref{SclA}) and (\ref{SclB}).

Next let us study the preexponential factor at two different magnetic field
directions of $\theta _H=0$ and $\pi /2\geqslant \theta _H>0.$

\begin{center}
{\bf I. }$\theta _H=0$
\end{center}

The preexponential factor can be evaluated from zero-mode $\left( d\phi
/d\tau \right) $. \cite{garg92} It is obvious from symmetry that the
preexponential factors along two instanton path $A$ and $B$ are equal.

For $\theta _H=0$, we have $\phi _0=0$ and $\phi _i=0,$ $\phi _f=\pi $.
Then, from Eq. (\ref{xy}) we get $x=\left( 1-\cos ^2\theta _0-\lambda \sin
^2\phi \right) ^{1/2}$ and $y=0$. Therefore, Eq. (\ref{fai1}) reduces to 
\begin{equation}
\dot{\phi}=\frac{2K_1}S\lambda ^{1/2}\left( 1-\cos ^2\theta _0-\lambda \sin
^2\phi \right) ^{1/2}\sin \phi .  \label{fai2}
\end{equation}
This equation can be integrated easily. Defining $\omega _0=\frac{2K_1}S%
\lambda ^{1/2}\sin \theta _0$, we obtain 
\begin{equation}
\cos \phi =\left( 1-\cos ^2\theta _0-\lambda \right) ^{1/2}\frac{\tanh
(\omega _0\tau )}{\left( 1-\cos ^2\theta _0-\lambda \tanh ^2\left( \omega
_0\tau \right) \right) ^{1/2}}.
\end{equation}
It is easily verified that $\phi \rightarrow 0,\pi $, as $\tau \rightarrow
\pm \infty $.

Following the standard procedure of the Ref. 10, we write the final result
as 
\begin{equation}
\hbar \Delta =c\omega _0^{3/2}\frac{8\sin \theta _0}{\left( 1-\cos ^2\theta
_0-\lambda \right) ^{1/2}}\left( \frac{S^2}{2\pi K_1}\right)
^{1/2}e^{-S_{cl}}\left| \cos \Phi (H)\right| ,  \label{asplit_a}
\end{equation}
where 
\begin{equation}
S_{cl}=2S\left\{ \frac 12\ln \left( \frac{1+\frac{\lambda ^{1/2}}{\sin
\theta _0}}{1-\frac{\lambda ^{1/2}}{\sin \theta _0}}\right) -\frac{\cos
\theta _0}{2\left( 1-\lambda \right) ^{1/2}}\ln \left( \frac{1+\frac{\lambda
^{1/2}\cos \theta _0}{\left( 1-\lambda \right) ^{1/2}\sin \theta _0}}{1-%
\frac{\lambda ^{1/2}\cos \theta _0}{\left( 1-\lambda \right) ^{1/2}\sin
\theta _0}}\right) \right\}  \label{Sr}
\end{equation}
and 
\begin{equation}
\Phi (H)=\pi S\left( 1-\frac{\cos \theta _0}{(1-\lambda )^{1/2}}\right) .
\label{phase}
\end{equation}
Eq. (\ref{asplit_a}) can be also deduced from Eqs. (\ref{SclA}), (\ref{ImDS}%
), (\ref{ReDS}) and (\ref{split0}) for the special case $\phi _0=0.$ It is
interesting to note that Eqs. (\ref{Sr}) and (\ref{phase}) agree exactly
with the previous result (Eqs. (3.10) and (3.11) in Ref. 7) found by Garg
for the molecular Fe$_8$ magnet.

\begin{center}
{\bf II. }$\pi /2\geqslant \theta _H>0$
\end{center}

\smallskip For $\pi /2\geqslant \theta _H>0$, it is difficult to integrate
the equation of motion Eq. (\ref{fai1}). So we have to take the asymptotic
form of zero-mode $\left( d\phi /d\tau \right) $. Notice that $\phi
\rightarrow \pi -\phi _0$ as $\tau \rightarrow +\infty $, we can expand $%
%TCIMACRO{\func{Im}}
%BeginExpansion
\mathop{\rm Im}%
%EndExpansion
f(\phi )$ according to the small parameter $\alpha =$ $\phi -(\pi -\phi _0)$%
. Up to the leading term, Eq. (\ref{fai1}) has the form 
\begin{equation}
\dot{\phi}=-\omega _0\left( \phi -(\pi -\phi _0)\right) ,  \label{fai3}
\end{equation}
where $\omega _0=\frac{2K_1}S\lambda ^{1/2}\sin \theta _0\cos \phi _0.$ We
then integrate the Eq. (\ref{fai3}) 
\begin{equation}
\phi =\pi -\phi _0-c_0e^{-\omega _0\tau },
\end{equation}
where $c_0$ is an integration constant. Therefore, the asymptotic form of
zero-mode $\left( d\phi /d\tau \right) $ can be written as 
\begin{equation}
d\phi /d\tau =c_0\omega _0e^{-\omega _0\tau }.  \label{solution_fai}
\end{equation}
Straightforward, following Ref. 10, the tunnel splitting $\hbar \Delta ^A$
and $\hbar \Delta ^B$ corresponding to path $A$ and $B$ have the form 
\begin{eqnarray}
\hbar \Delta ^A &=&c_A\omega _0^{3/2}\left( \frac{S^2}{2\pi K_1}\right)
^{1/2}\left( \frac{\sin ^2\theta _0}{\sin ^2\theta _0+\lambda \cos ^2\theta
_0\sin ^2\phi _0}\right) ^{1/2}e^{-%
%TCIMACRO{\func{Re}}
%BeginExpansion
\mathop{\rm Re}%
%EndExpansion
(S_{cl}^A)}e^{-i%
%TCIMACRO{\func{Im}}
%BeginExpansion
\mathop{\rm Im}%
%EndExpansion
(S_{cl}^A)}  \label{asplit_b1} \\
&&  \nonumber \\
\hbar \Delta ^B &=&c_B\omega _0^{3/2}\left( \frac{S^2}{2\pi K_1}\right)
^{1/2}\left( \frac{\sin ^2\theta _0}{\sin ^2\theta _0+\lambda \cos ^2\theta
_0\sin ^2\phi _0}\right) ^{1/2}e^{-%
%TCIMACRO{\func{Re}}
%BeginExpansion
\mathop{\rm Re}%
%EndExpansion
(S_{cl}^B)}e^{-i%
%TCIMACRO{\func{Im}}
%BeginExpansion
\mathop{\rm Im}%
%EndExpansion
(S_{cl}^B)},  \label{asplit_b2}
\end{eqnarray}
where $c_A$ and $c_B$ are the numerical factors in order of $O(1)$ that come
from the integration constant mentioned above ($c_0$ in Eq. (\ref
{solution_fai})). Because the existence of $H_y$ breaks the symmetry between
the classical path $A$ and $B$, $c_A$ may not be equal to $c_B$. But for the
small magnetic field, we assume theirs equivalence. Comparing Eqs. (\ref
{asplit_b1}) and (\ref{asplit_b2}) with Eq. (\ref{asplit_a}), we find 
\begin{equation}
c_A=c_B=c\frac{4\sin \theta _0}{\left( 1-\cos ^2\theta _0-\lambda \right)
^{1/2}}.
\end{equation}

Now we turn to derive the total tunnel splitting $\hbar \Delta $ using a
recently proposed effective Hamiltonian approach. For the present case, the
effective Hamiltonian is found to be 
\begin{equation}
{\cal H}_{eff}=\left[ 
\begin{array}{c}
0 \\ 
\left( \hbar \Delta ^A+\hbar \Delta ^B\right) ^{*}
\end{array}
\begin{array}{c}
\left( \hbar \Delta ^A+\hbar \Delta ^B\right) \\ 
0
\end{array}
\right] .
\end{equation}
Diagonalizing the effective Hamiltonian, we obtain the desired result as
shown in Eqs. (\ref{split0}) and (\ref{split1}).

\newpage 
\begin{tabular}{ccccccccc}
\hline\hline
&  & 
\begin{tabular}{c}
$\theta _H=0^o$ \\ \hline
$\Delta _d\quad \qquad \Delta _i$%
\end{tabular}
&  & 
\begin{tabular}{c}
$\theta _H=15^o$ \\ \hline
$\Delta _d\qquad \quad \Delta _i$%
\end{tabular}
&  & 
\begin{tabular}{c}
$\theta _H=30^o$ \\ \hline
$\Delta _d\quad \qquad \Delta _i$%
\end{tabular}
&  & 
\begin{tabular}{c}
$\theta _H=90^o$ \\ \hline
$\Delta _d\quad \qquad \Delta _i$%
\end{tabular}
\\ \hline
$H_S=0.0$T & \qquad & 0.00683\quad 0.00683 & \qquad & 0.00683\quad 0.00683 & 
\qquad & 0.00683\quad 0.00683 & \qquad & 0.00683\quad 0.00683 \\ 
$H_S=0.2$T &  & 0.00556\quad 0.00558 &  & 0.00778\quad 0.00780 &  & 
0.01445\quad 0.01448 &  & 0.05651\quad 0.05658 \\ 
$H_S=0.4$T &  & 0.00165\quad 0.00167 &  & 0.01714\quad 0.01733 &  & 
0.06521\quad 0.06591 &  & 0.75350\quad 0.75752 \\ 
$H_S=0.6$T &  & 0.00502\quad 0.00515 &  & 0.04658\quad 0.04782 &  & 
0.29419\quad 0.30232 &  & 8.20514\quad 8.30553 \\ 
$H_S=0.8$T &  & 0.01417\quad 0.01485 &  & 0.13101\quad 0.13783 &  & 
1.34388\quad 1.42229 &  & 73.1429\quad 74.7913 \\ 
$H_S=1.0$T &  & 0.02362\quad 0.02548 &  & 0.39753\quad 0.43399 &  & 
6.19230\quad 6.88449 &  & 537.366\quad 557.203 \\ 
$H_S=1.2$T &  & 0.02505\quad 0.02808 &  & 1.31263\quad 1.51459 &  & 
28.5282\quad 34.3140 &  & 3277.23\quad 3462.35 \\ 
$H_S=1.4$T &  & 0.00799\quad 0.00942 &  & 4.70886\quad 5.90363 &  & 
129.907\quad 176.358 &  & 16704.7\quad 18092.3 \\ \hline\hline
\end{tabular}

\begin{center}
\bigskip

Table 1. The ground-state tunnel splitting $\Delta _d$ (calculated by
diagonalization) and $\Delta _i$ (calculated by instanton approach) with $%
\theta _H=0^o$, $15^o$, $30^o$ and $90^o$ are listed as a function of
magnetic field $H_S$. The units for the tunnel splitting and magnetic field
are $10^{-7}$ Kelvin and Tesla, respectively. Here, $S=10$, $K_1=32.1$K, $%
K_2=22.9$K and $\lambda =0.71$ are used for the molecular Fe$_8$ magnet.

\newpage

{\bf Figures Captions}
\end{center}

Fig. 1. The critical magnetic field $H_c$ is plotted as a function of anlge $%
\theta _H$ at $\lambda =0.71$.

Fig. 2. The ground-state tunnel splitting $\Delta $ are plotted as a
function of the magnetic field $H_S$ at $\theta _H=0^o,$ $3^o,$ $10^o,$ $%
30^o,$ $60^o,$ and $90^o$ for (a) $S=10$ and (b) $S=9.5$. The other
parameters are $K_1=32.1$K, $K_2=22.9$K and $\lambda =0.71,$ the typical
values of the molecular Fe$_8$ magnet. The results of the instanton approach
and the exact diagonalization are represented by the solid lines and square
symbols, respectively.

Fig. 3. The ground-state tunnel splitting $\Delta $ are plotted as a
function of the magnetic field $H_S$ for $\lambda =0.3,$ $0.5$ and $0.7$
with $S=10$, $K_1=32.1$K at (a) $\theta _H=0^o$ and (b) $\theta _H=90^o$.
The results of the instanton approach and the exact diagonalization are
represented by the solid lines and square symbols, respectively.


\begin{references}
\bibitem{overview}  For an overview, see e.g., {\it Quantum Tunneling of
Magnetization - QTM'94}, edited by L. Gunther and B. Barbara (Kluwer,
Dordrecht, The Netherlands, 1994), and references therein.

\bibitem{Mn12Ac}  R. Sessoli, D. Gatteschi, A. Caneschi, and M. A. Novak,
Nature {\bf 365}, 141 (1993).

D. Gatteschi, A. Caneschi, L. Pardi, and R. Sessoli, Science {\bf 265}, 1054
(1994).

L. Thomas, F. Lionti, R. Ballou, D. Gatteschi, R. Sessoli, and B. Barbara,
Nature {\bf 383}, 145 (1996).

J. M. Hernandez, X. X. Zhang, F. Luis, J. Tejada, J. R. Friedman, M. P.
Sarachik, and R. Ziolo, Phys. Rev. B {\bf 55}, 5858 (1997).

D. A. Garanin and E. M. Chudnovsky, Phys. Rev. B {\bf 56}, 11102(1997).

\bibitem{Fe8}  A. -L. Barra, P. Debrunner, A. Catteschi, C. E. Schulz, and
R. Sessoli, Europhys. Lett. {\bf 35}, 133 (1996).

C. Sangregorio, T. Ohm, C. Paulsen, R. Sessoli, and D. Gatteschi, Phys. Rev.
Lett. {\bf 78}, 4645 (1997).

\bibitem{loss}  D. Loss, D. P. Divicenzo, and G. Grinstein, Phys. Rev. Lett. 
{\bf 69}, 3232 (1992).

\bibitem{delft}  J. V. Delft and G. L. Henley, Phys. Rev. Lett. {\bf 69},
3236 (1992).

\bibitem{garg93}  A. Garg, Europhys. Lett. {\bf 22}, 205 (1993).

\bibitem{garg99}  A. Garg, Phys. Rev. B {\bf 60}, 6705 (1999).

\bibitem{braun}  H. B. Braun and D. Loss, Europhys. Lett. {\bf 31}, 555
(1995).

\bibitem{wernsd}  W. Wernsdorfer and R. Sessoli, Science {\bf 284}, 133
(1999).

\bibitem{garg92}  A. Garg and G. -H. Kim, Phys. Rev. B {\bf 45}, 12921
(1992).
\end{references}
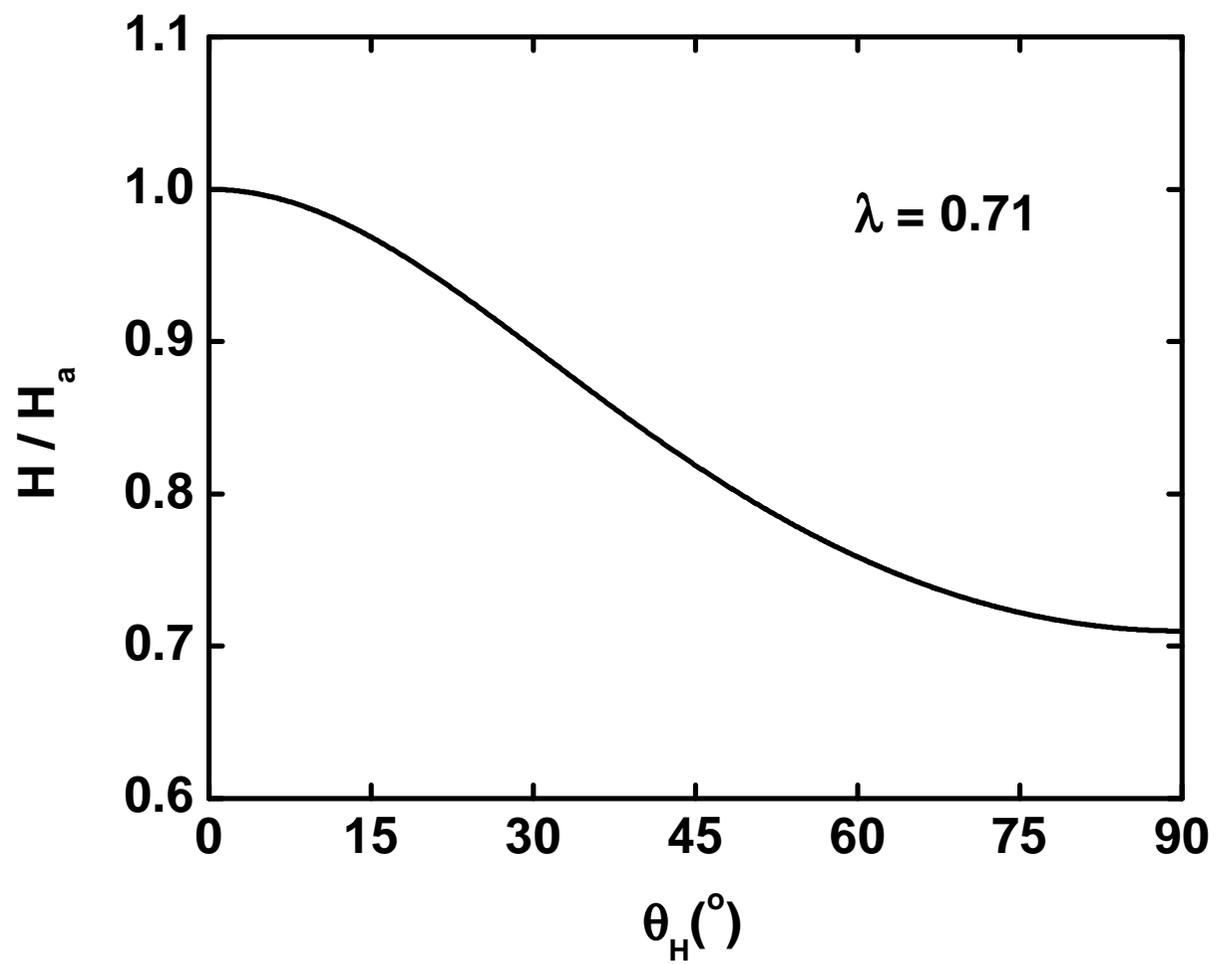
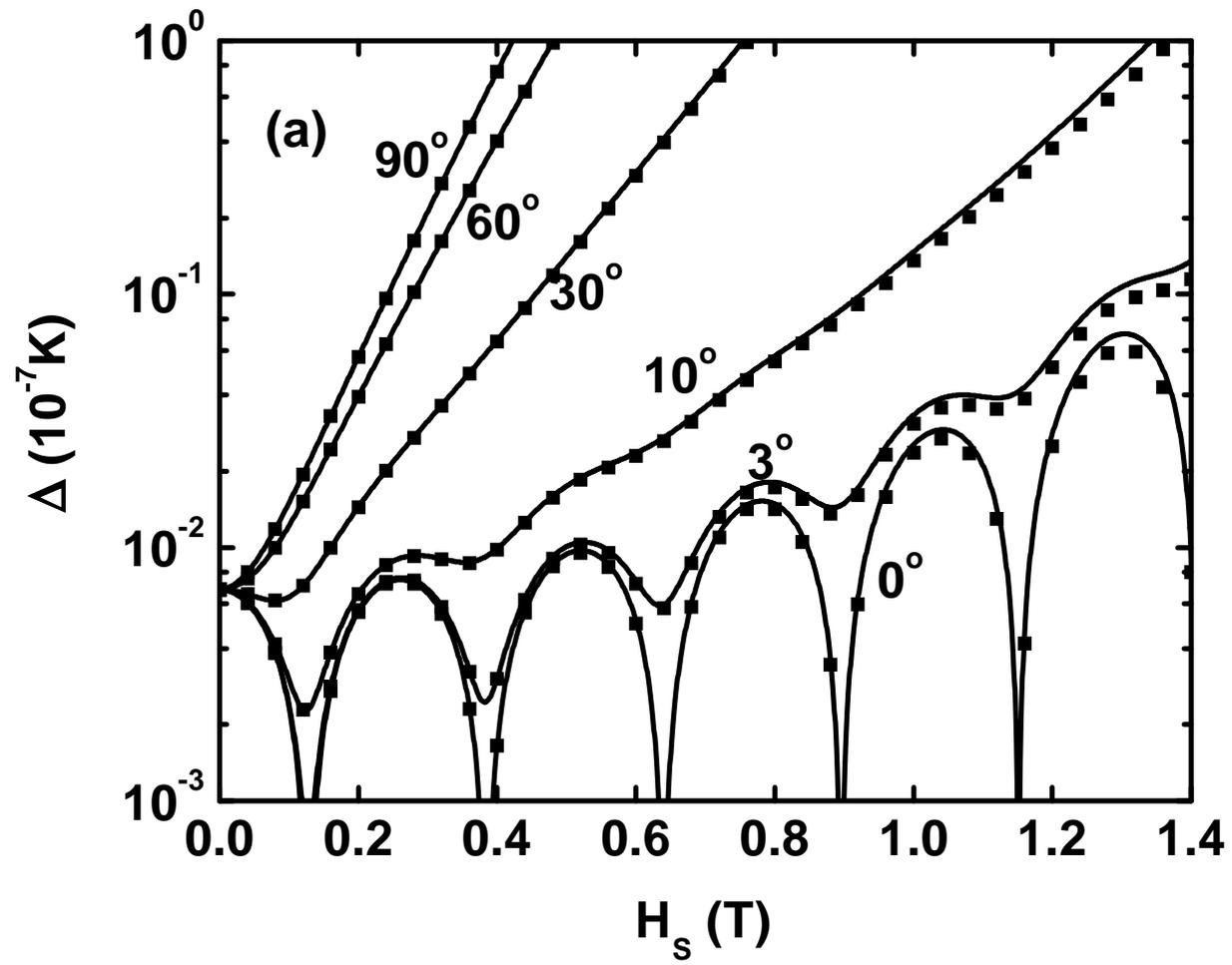
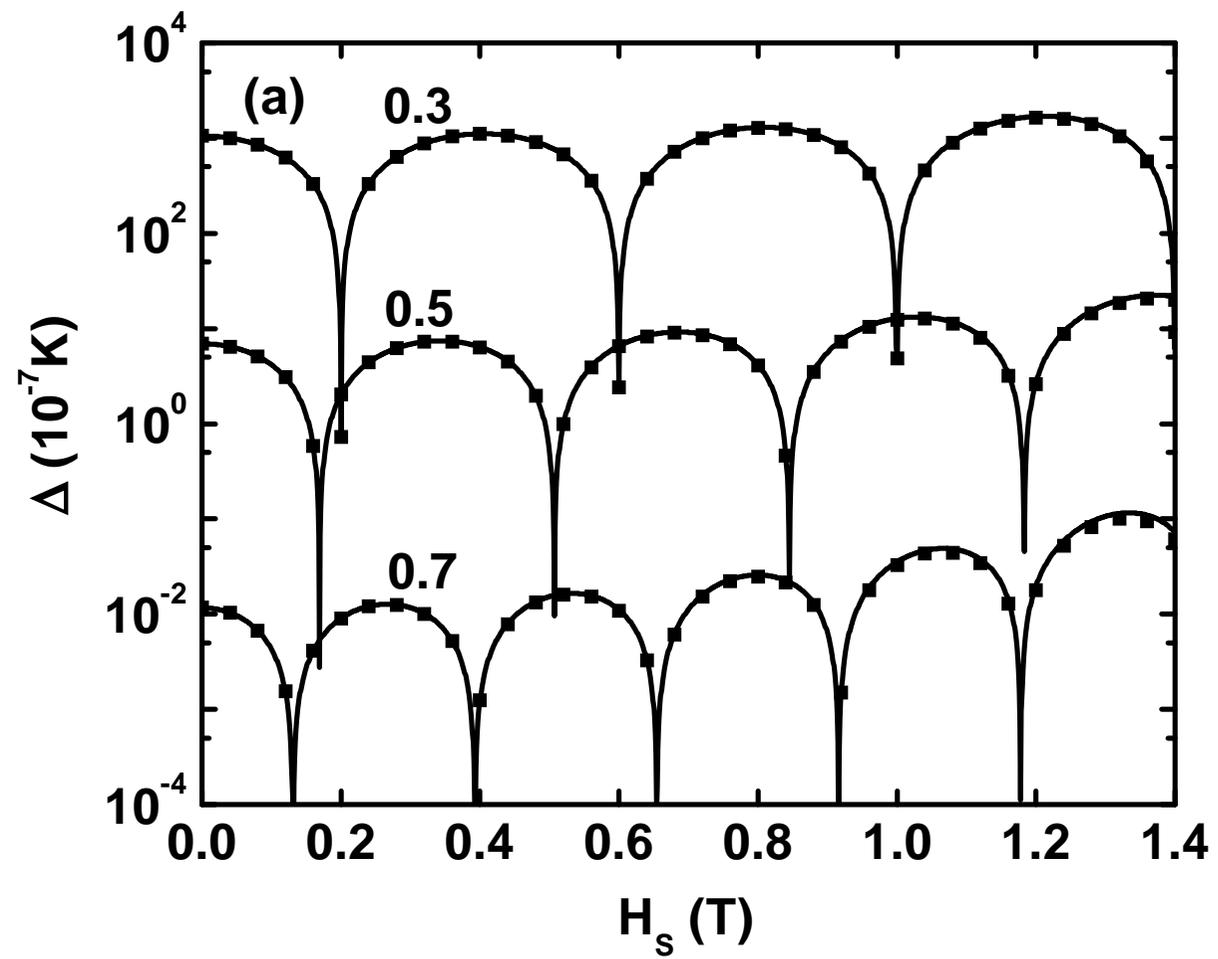
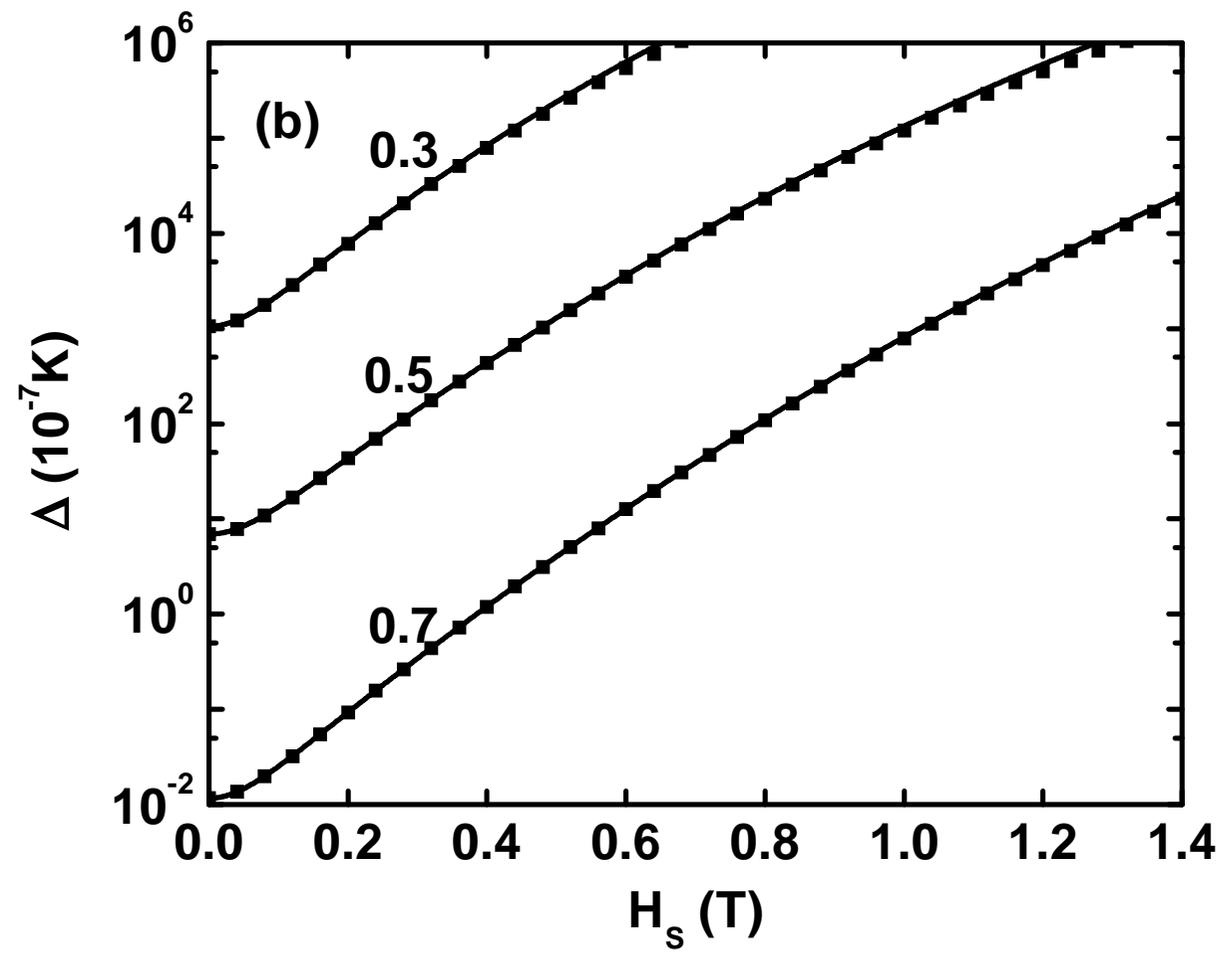
\end{document}